\documentclass[jmp,twocolumn,showpacs,preprintnumbers,amsmath,amssymb]{revtex4-1}
\usepackage{simplewick}

\usepackage{graphicx}
\usepackage{dcolumn}
\usepackage{bm}
\newcommand \be{\begin{eqnarray}}
\newcommand \ee{\end{eqnarray}}
\newcommand \ba{\begin{align}}
\newcommand \eea{\end{align}}

\begin{document}
           \csname @twocolumnfalse\endcsname
\title{Charged hanging chain - catenary exposed to force}
\author{K. Morawetz$^{1,2}$}
\affiliation{$^1$M\"unster University of Applied Science,
Stegerwaldstrasse 39, 48565 Steinfurt, Germany}
\affiliation{$^2$International Institute of Physics (IIP)
Federal University of Rio Grande do Norte
Av. Odilon Gomes de Lima 1722, 59078-400 Natal, Brazil
}
\begin{abstract}
A new solution of a charged catenary is presented which allows to determine the static stability conditions where charged
liquid bridges or charged hanging chains are possible.
\end{abstract}
\maketitle

The problem of hanging chains forming a catenary is one of the standard text book examples in variation theory to minimize the free energy. The typical applications ranging from elastic chains, cables, up to liquid bridge formations.
The formation of a water bridge between two beakers when high electric fields
are applied is a phenomenon known since over 100 years \cite{arm1893} and has
remained attractive to current experimental activities \cite{Wo10,ML10}. Such formation of water bridges on nanoscales are of current
interest both for fundamental understanding of electrohydrodynamics  and for 
applications ranging from atomic force microscopy \cite{Sa06} to electro-wetting problems \cite{Ju10}. Microscopically the nanoscale wetting is important
to confine chemical reactions \cite{GM06} which reveals an interesting
interplay between field-induced polarization, surface tension, and
condensation \cite{GS03,CZG08}. In this context is might be interesting to note that an analytic solution of the bulk-charged catenary is possible.

In the absence of bulk charges the forces on the liquid stream are caused by the
pressure due to the polarizability of the liquid described by the high dielectric susceptibility
$\epsilon$. This pressure leads to the catenary form of water bridge 
like a hanging chain \cite{WSSSS09}. While already the simplified model of \cite{Sa10} employing a capacitor picture leads to a critical field strength for the formation of
the water bridge, the catenary model \cite{WSSSS09} has not been reported to
yield such a
critical field. In this paper we will show that even the uncharged catenary
provides indeed a minimal critical field strength for the water bridge
formation in dependence on the length of the bridge. This critical field
strength is modified if charges are present in the bridge which we will
discuss here with the help of a new charged catenary solution. This allows us to explain the asymmetry found in the bridge profile \cite{ML10}.

Imagine a charged chain or a liquid bridge carrying bulk charges under the influence of an electric field. The form of the catenary will certainly be deformed due to the field. We will describe the analytic solution in terms of two parameters.
One is the liquid column height balancing the dielectric
pressure called creeping height in the following
\be
b(E)={\epsilon_0(\epsilon-1) E^2\over  \rho g}
\label{b}
\ee
where $\epsilon$ is the dielectric constant of the medium, $\rho$ the number density $g$ the gravity and $E$ the applied electric field. For a charged chain this would be the height where the gravitation energy becomes equal to the pressure exercised by the field. Further we will need the dimensionless ratio of the force density on the
charges by the field to the gravitational force density
\be
c(\rho_c,E)={\rho_c E\over \rho g}.
\label{c}
\ee 
where the charge density is ${\rho_c}$.

We consider the center-of-mass line of the bridge or chain being described by $z=f(x)$ with the ends at $f(0)=f(L)=0$. The force densities are multiplied with the area and the length element $ds=\sqrt{1+f'^2} dx$ to  form the free energy. We have the gravitational force density$ \rho g f$ and the volume tension $\rho g b$ as well as the force density by dynamical charges $\rho_c E x$ which contributes. The surface tension is negligible here. The form of the bridge or chain will be then determined by the extreme value of the free energy
\be
&&\int\limits_0^L {\cal F}(x) dx=\rho g \int\limits_0^L (f(x)+b-c x) \sqrt{1+f'^2} dx \to {\rm extr.}\nonumber\\&&
\label{extr}
\ee
where $c$ is given by (\ref{c}) and $b$ defined in (\ref{b}) and the boundary conditions $f(0)=f(L)=0$. It is useful to introduce
\be
t(x)= f(x)+b-c x
\ee
such that
\be
{\cal F}(x)=\rho g \,  t(x) \sqrt{1+[t'(x)+c]^2}.
\ee
The corresponding Lagrange equation 
\be
{d\over d x} {\partial {\cal F}\over \partial t'(x)}-{\partial {\cal F}\over \partial t(x)}=0
\ee
possesses a first
integral
\be
t'(x) {\partial {\cal F}\over \partial t'(x)}-{\cal F}={\rm const}=-\xi \sqrt{1+c^2}
\ee
where we introduced the first integration constant $\xi$ in a convenient way. 
The resulting differential equation 
\be
t(\bar x) [c t'(\bar x)+1]=\xi \sqrt{t'(\bar x)^2+(c t'(\bar x)+1)^2}
\ee
with $\bar x =x (1+c^2)$
is solved in an implicit way
\be
t(\bar x)=\xi \cosh\left \{\frac 1 \xi \left [\bar x+c t(\bar x)-c b+\frac L 2 d\right ]\right \}
\label{f1}
\ee
with a second integration constants $d$ which is determined by the boundary condition $f(0)=0$ as
\be
d=2 {\xi \over L}{\rm arcosh}\left (b\over \xi \right )
\label{dl}
\ee
in terms of the yet unknown $\xi$ constant. Eq. (\ref{f1}) can be written with the help of (\ref{dl}) as
\ba
f(x)=c x \!+\!\xi\left \{ \!\cosh\!\!\left [{ x\!+\!cf(x)\over \xi} \!-\!{L d\over 2 \xi} \right] \!-\!\cosh\!\left (\!{L d\over 2 \xi}\! \right )\!\right \}.
\label{f2}
\end{align}
The boundary condition $f(L)=0$ leads then to the determination of the remaining constant $\xi$ as solution of the equation
\be
c&=&c_m(\xi,b)\nonumber\\
c_m(\xi,b)\!&=&\!-{2 \xi \over L}\sinh{L\over 2\xi} \!\left ( {b\over \xi} \sinh{L\over 2 \xi}\!-\!\sqrt{{b^2\over \xi^2}-1} \cosh{L\over 2 \xi}\right ).\nonumber\\&&
\label{fa}
\ee
Finally we choose as parameter $t=x+c f(x)$ which runs obviously through the interval $t\in (0,L)$ and we obtain from (\ref{f2}) the parametric representation of the solution
\be
f(t)&=&{1\over 1\!+\!c^2}\left \{c \,t\!+\!\xi \left [\cosh{\left (\frac t
      \xi\!-\!\frac{ L d}{2\xi}\right )}\!-\!\cosh{\left (L  d\over 2\xi\right
    )}\right ]\right \}\nonumber\\
x(t)&=&t-c f(t),\qquad t\in(0,L).
\label{sol}
\ee
with $d$ and $\xi$ obeying the equations (\ref{dl}) and
 (\ref{fa}).

\begin{figure}
\includegraphics[width=8cm]{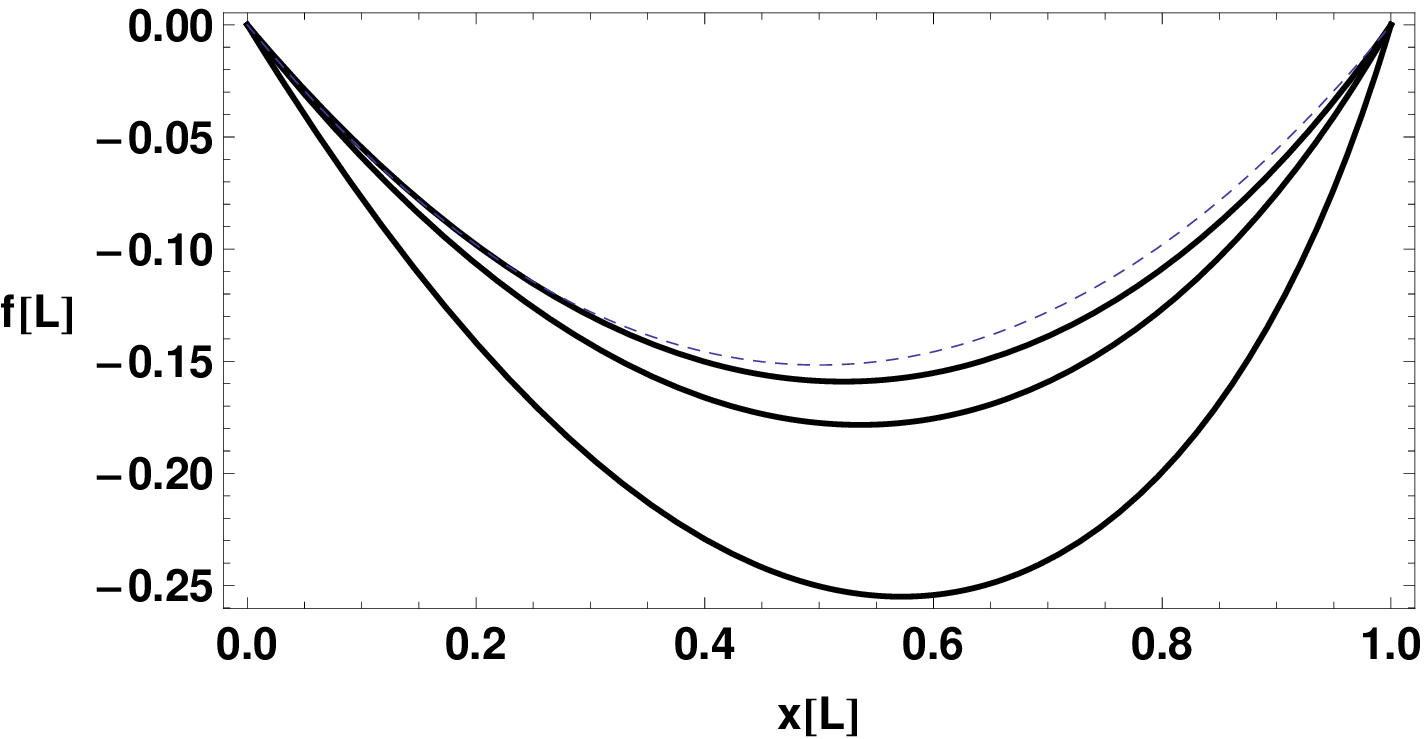}
\includegraphics[width=8cm]{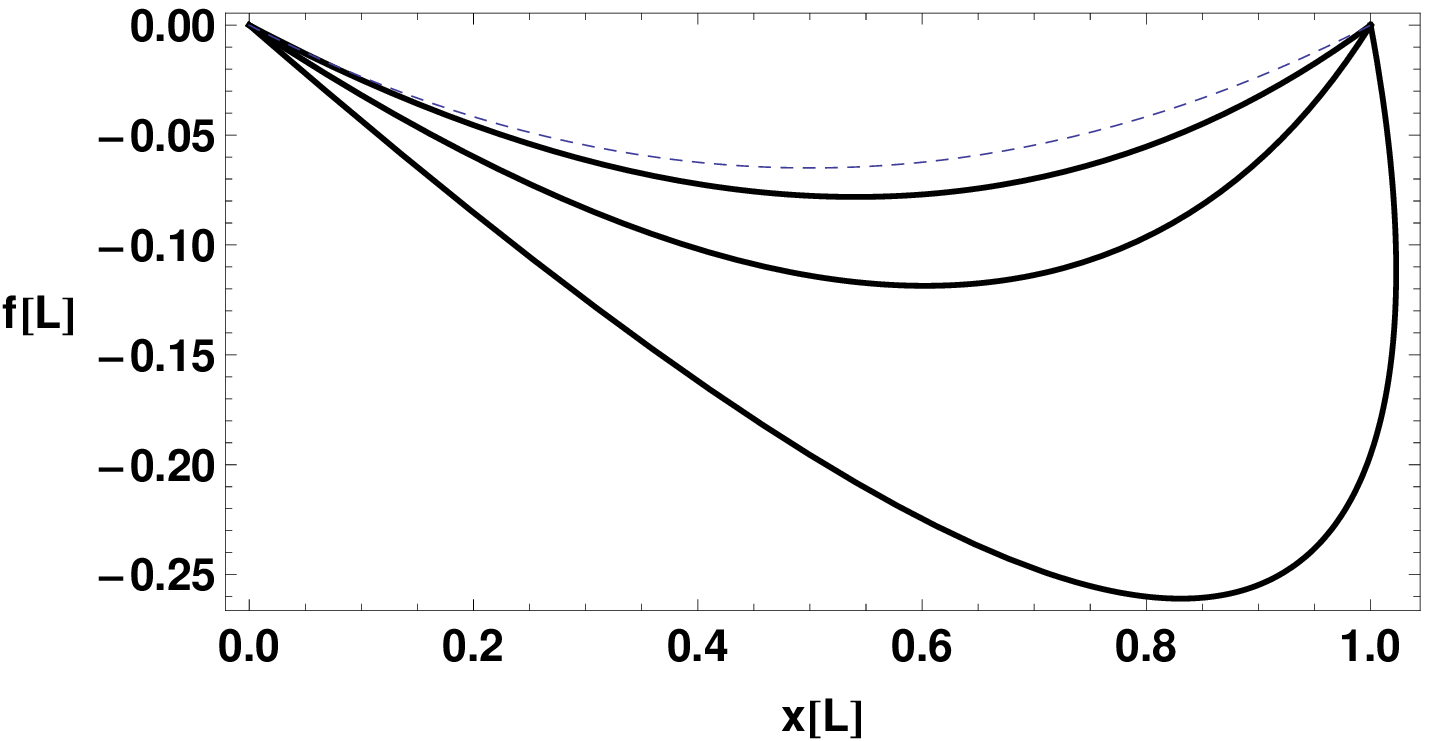}
\includegraphics[width=8cm]{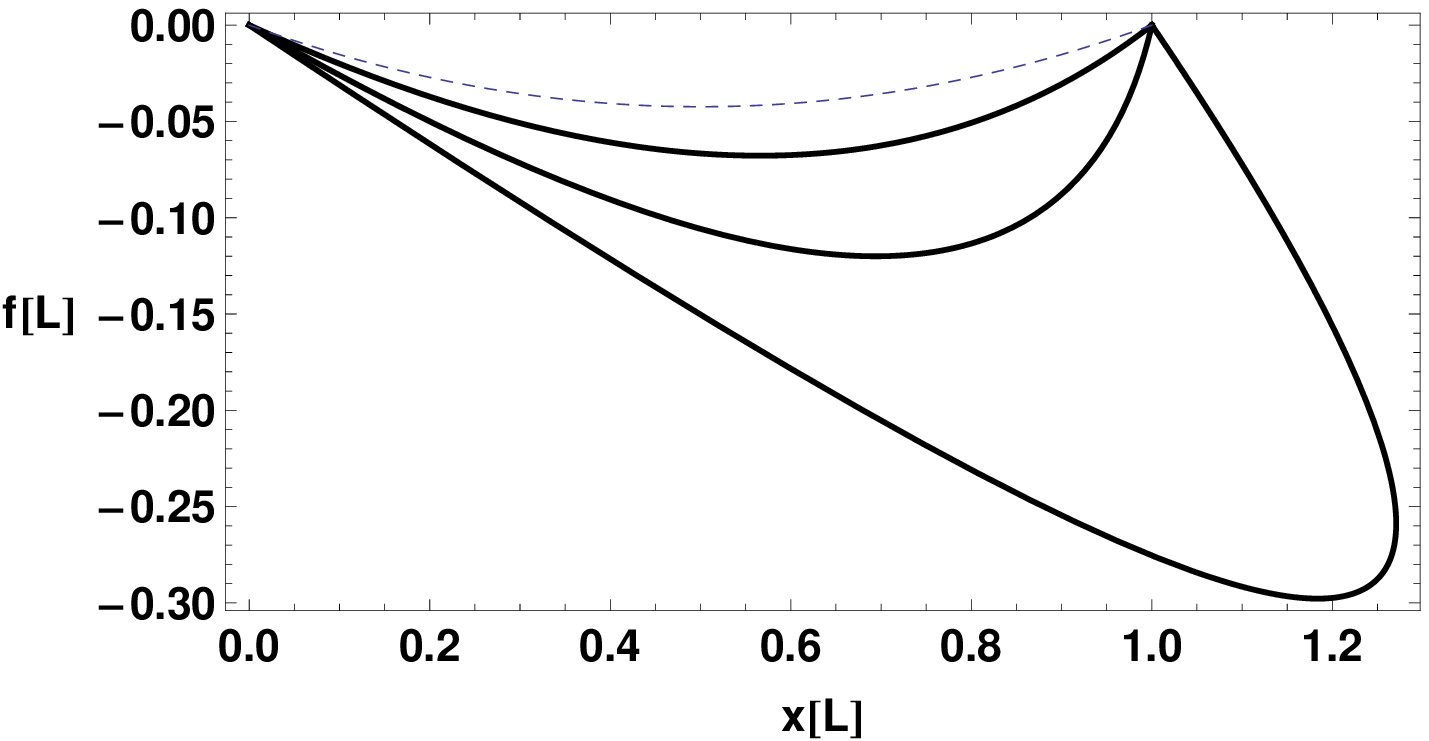}
\caption{\label{catb} The charged catenary profile (\ref{sol}) for $b=1,
2,
3$ (from above to below) corresponding to the maximal values $c_m=0.41, 
1.62, 
2.68$. The uncharged catenary is given by dashed lines and the three thick lines are $c=c_m (0.5,0.75,0.999)$ respectively.}
\end{figure}

The charged catenary profile (\ref{sol}) has not been reported in the literature so far and is
the main result of this letter. The profiles are plotted in figure \ref{catb} in comparison to the uncharged catenary. One sees that the deformation due to the electric field in x-direction can lead to appreciable inelastic deformations which parameter are artificially chosen here. Such large deformations are expected for charged elastic materials perhaps but certainly not for water bridges.

Without dynamical bulk charges, $c=0, d=1$, the solution (\ref{sol}) is just the well known catenary \cite{WSSSS09}. The boundary condition (\ref{fa}) reads in this case
\be
{2 b\over L}={2 \xi \over L}\cosh{L\over 2 \xi}\ge \xi_c=1.5088...
\ee
which means that without bulk charges the condition for a stable bridge is
\be
b > \frac L 2 \xi_c.
\label{cond1a}
\ee 
Together with (\ref{b}) this condition provides a lower bound for the electric field
in order to enable a bridge of length $L$. This lower bound for an applied field  appears obviously already for the standard catenary  and has been not discussed so far.
 
\begin{figure}
\includegraphics[width=9cm]{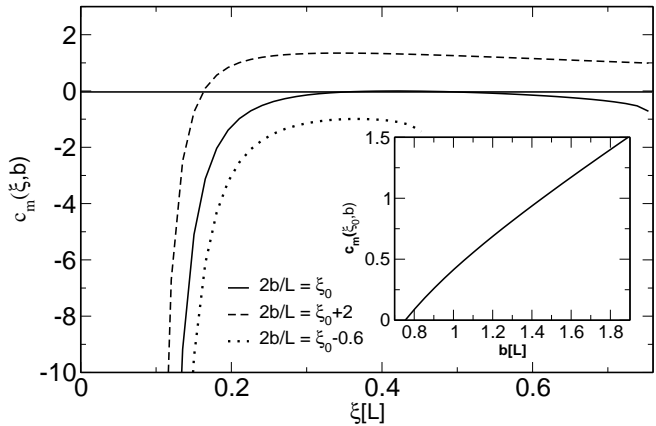}
\caption{\label{figure1} The upper critical bound for the parameter $c$
  according to (\ref{fa}). The inset shows the maximum in dependence on the
  creeping parameter $b$.}
\end{figure}

Lets now return to the more involved case of bulk charges and the new solution of charged catenary (\ref{sol}).
The field-dependent lower bound  condition (\ref{fa}) is plotted in
figure \ref{figure1}. One see that in order to complete (\ref{fa}) the bulk
charge parameter $c$ has to be lower than the maximal value of $c_m$ which reads
\be
c\le c_{max}(\xi_0,b)
\label{cond1}
\ee
and which is plotted in the inset of figure \ref{figure1}. Remembering the
definition of the bulk charge parameter (\ref{c}) we see that (\ref{cond1})
sets an upper bound for the bulk charge in dependence on the electric field.
The lower bound (\ref{cond1a}) of the electric field 
for the case of no bulk charges is obeyed as well since the curve in the inset
of figure \ref{figure1} starts at $b>L \xi_c/2$ which is the lower bound already present for uncharged catenaries (\ref{cond1a}). 

This completes the question concerning static stability of the
bridge or chain in an electric field. We have found a new catenary solution for bulk charges in the liquid bridges or charged chains. The corresponding application might be seen in elastic deformations of chains due to charges in an electric field or in deformations of charged liquid bridges.

This work was supported by DFG project within SPP 1158. The financial support by the Brazilian Ministry of Science 
and Technology is acknowledged.


\begin{thebibliography}{10}%
\makeatletter
\providecommand \@ifxundefined [1]{%
 \@ifx{#1\undefined}
}%
\providecommand \@ifnum [1]{%
 \ifnum #1\expandafter \@firstoftwo
 \else \expandafter \@secondoftwo
 \fi
}%
\providecommand \@ifx [1]{%
 \ifx #1\expandafter \@firstoftwo
 \else \expandafter \@secondoftwo
 \fi
}%
\providecommand \natexlab [1]{#1}%
\providecommand \enquote  [1]{``#1''}%
\providecommand \bibnamefont  [1]{#1}%
\providecommand \bibfnamefont [1]{#1}%
\providecommand \citenamefont [1]{#1}%
\providecommand \href@noop [0]{\@secondoftwo}%
\providecommand \href [0]{\begingroup \@sanitize@url \@href}%
\providecommand \@href[1]{\@@startlink{#1}\@@href}%
\providecommand \@@href[1]{\endgroup#1\@@endlink}%
\providecommand \@sanitize@url [0]{\catcode `\\12\catcode `\$12\catcode
  `\&12\catcode `\#12\catcode `\^12\catcode `\_12\catcode `\%12\relax}%
\providecommand \@@startlink[1]{}%
\providecommand \@@endlink[0]{}%
\providecommand \url  [0]{\begingroup\@sanitize@url \@url }%
\providecommand \@url [1]{\endgroup\@href {#1}{\urlprefix }}%
\providecommand \urlprefix  [0]{URL }%
\providecommand \Eprint [0]{\href }%
\@ifxundefined \urlstyle {%
  \providecommand \doi  [0]{\begingroup \@sanitize@url \@doi}%
  \providecommand \@doi [1]{\endgroup \@@startlink {\doibase
  #1}doi:\discretionary {}{}{}#1\@@endlink }%
}{%
  \providecommand \doi  [0]{doi:\discretionary{}{}{}\begingroup
  \urlstyle{rm}\Url }%
}%
\providecommand \doibase [0]{http://dx.doi.org/}%
\providecommand \Doi [0]{\begingroup \@sanitize@url \@Doi }%
\providecommand \@Doi  [1]{\endgroup\@@startlink{\doibase#1}\@@Doi}%
\providecommand \@@Doi [1]{#1\@@endlink}%
\providecommand \selectlanguage [0]{\@gobble}%
\providecommand \bibinfo  [0]{\@secondoftwo}%
\providecommand \bibfield  [0]{\@secondoftwo}%
\providecommand \translation [1]{[#1]}%
\providecommand \BibitemOpen [0]{}%
\providecommand \bibitemStop [0]{}%
\providecommand \bibitemNoStop [0]{.\EOS\space}%
\providecommand \EOS [0]{\spacefactor3000\relax}%
\providecommand \BibitemShut  [1]{\csname bibitem#1\endcsname}%
\bibitem [{\citenamefont {Armstrong}(1893)}]{arm1893}%
  \BibitemOpen
  \bibfield  {author} {\bibinfo {author} {\bibfnamefont {W.~G.}\ \bibnamefont
  {Armstrong}},\ }\enquote {\bibinfo {title} {The electrical engineer},}\ \
  (\bibinfo  {publisher} {The Newcastle Literary and Philosophical Society},\
  \bibinfo {address} {New Castle},\ \bibinfo {year} {1893})\ pp.\ \bibinfo
  {pages} {154--155},\ \bibinfo {note} {18 February 1893}\BibitemShut {NoStop}%
\bibitem [{\citenamefont {Woisetschlager}\ \emph {et~al.}(2010)\citenamefont
  {Woisetschlager}, \citenamefont {Gatterer},\ and\ \citenamefont
  {Fuchs}}]{Wo10}%
  \BibitemOpen
  \bibfield  {author} {\bibinfo {author} {\bibfnamefont {J.}~\bibnamefont
  {Woisetschlager}}, \bibinfo {author} {\bibfnamefont {K.}~\bibnamefont
  {Gatterer}}, \ and\ \bibinfo {author} {\bibfnamefont {E.}~\bibnamefont
  {Fuchs}},\ }\href@noop {} {\bibfield  {journal} {\bibinfo  {journal} {Exp. in
  Fluids},\ }\textbf {\bibinfo {volume} {48}},\ \bibinfo {pages} {121 }
  (\bibinfo {year} {2010})}\BibitemShut {NoStop}%
\bibitem [{\citenamefont {Marin}\ and\ \citenamefont {Lose}(2010)}]{ML10}%
  \BibitemOpen
  \bibfield  {author} {\bibinfo {author} {\bibfnamefont {A.~G.}\ \bibnamefont
  {Marin}}\ and\ \bibinfo {author} {\bibfnamefont {D.}~\bibnamefont {Lose}},\
  }\href@noop {} {\bibfield  {journal} {\bibinfo  {journal} {Phys. of Fluids},\
  }\textbf {\bibinfo {volume} {22}},\ \bibinfo {pages} {122104} (\bibinfo
  {year} {2010})}\BibitemShut {NoStop}%
\bibitem [{\citenamefont {Sacha}\ \emph {et~al.}(2006)\citenamefont {Sacha},
  \citenamefont {Verdaguer},\ and\ \citenamefont {Salmeron}}]{Sa06}%
  \BibitemOpen
  \bibfield  {author} {\bibinfo {author} {\bibfnamefont {G.}~\bibnamefont
  {Sacha}}, \bibinfo {author} {\bibfnamefont {A.}~\bibnamefont {Verdaguer}}, \
  and\ \bibinfo {author} {\bibfnamefont {M.}~\bibnamefont {Salmeron}},\
  }\href@noop {} {\bibfield  {journal} {\bibinfo  {journal} {J. Phys. Chem.
  B},\ }\textbf {\bibinfo {volume} {110}},\ \bibinfo {pages} {14870 } (\bibinfo
  {year} {2006})}\BibitemShut {NoStop}%
\bibitem [{\citenamefont {Oh}\ \emph {et~al.}(2010)\citenamefont {Oh},
  \citenamefont {Ko},\ and\ \citenamefont {Kang}}]{Ju10}%
  \BibitemOpen
  \bibfield  {author} {\bibinfo {author} {\bibfnamefont {J.~M.}\ \bibnamefont
  {Oh}}, \bibinfo {author} {\bibfnamefont {S.~H.}\ \bibnamefont {Ko}}, \ and\
  \bibinfo {author} {\bibfnamefont {K.~H.}\ \bibnamefont {Kang}},\ }\href@noop
  {} {\bibfield  {journal} {\bibinfo  {journal} {Physics of Fluids},\ \bibinfo
  {pages} {032002}} (\bibinfo {year} {2010})}\BibitemShut {NoStop}%
\bibitem [{\citenamefont {Garcia-Martin}\ and\ \citenamefont
  {Garcia}(2006)}]{GM06}%
  \BibitemOpen
  \bibfield  {author} {\bibinfo {author} {\bibfnamefont {A.}~\bibnamefont
  {Garcia-Martin}}\ and\ \bibinfo {author} {\bibfnamefont {R.}~\bibnamefont
  {Garcia}},\ }\href@noop {} {\bibfield  {journal} {\bibinfo  {journal} {Appl.
  Phys. Lett.},\ }\textbf {\bibinfo {volume} {88}},\ \bibinfo {pages} {123115}
  (\bibinfo {year} {2006})}\BibitemShut {NoStop}%
\bibitem [{\citenamefont {Gomez-Monivas}\ \emph {et~al.}(2003)\citenamefont
  {Gomez-Monivas}, \citenamefont {Saenz}, \citenamefont {Calleja},\ and\
  \citenamefont {Garcia}}]{GS03}%
  \BibitemOpen
  \bibfield  {author} {\bibinfo {author} {\bibfnamefont {S.}~\bibnamefont
  {Gomez-Monivas}}, \bibinfo {author} {\bibfnamefont {J.}~\bibnamefont
  {Saenz}}, \bibinfo {author} {\bibfnamefont {M.}~\bibnamefont {Calleja}}, \
  and\ \bibinfo {author} {\bibfnamefont {R.}~\bibnamefont {Garcia}},\
  }\href@noop {} {\bibfield  {journal} {\bibinfo  {journal} {Phys. Rev.
  Lett.},\ }\textbf {\bibinfo {volume} {91}} (\bibinfo {year}
  {2003})}\BibitemShut {NoStop}%
\bibitem [{\citenamefont {Cramer}\ \emph {et~al.}(2008)\citenamefont {Cramer},
  \citenamefont {Zerbetto},\ and\ \citenamefont {Garcia}}]{CZG08}%
  \BibitemOpen
  \bibfield  {author} {\bibinfo {author} {\bibfnamefont {T.}~\bibnamefont
  {Cramer}}, \bibinfo {author} {\bibfnamefont {F.}~\bibnamefont {Zerbetto}}, \
  and\ \bibinfo {author} {\bibfnamefont {R.}~\bibnamefont {Garcia}},\
  }\href@noop {} {\bibfield  {journal} {\bibinfo  {journal} {Langmuir},\
  }\textbf {\bibinfo {volume} {24}},\ \bibinfo {pages} {6116 } (\bibinfo {year}
  {2008})}\BibitemShut {NoStop}%
\bibitem [{\citenamefont {Widom}\ \emph {et~al.}(2009)\citenamefont {Widom},
  \citenamefont {Swain}, \citenamefont {Silverberg}, \citenamefont
  {Sivasubramanian},\ and\ \citenamefont {Srivastava}}]{WSSSS09}%
  \BibitemOpen
  \bibfield  {author} {\bibinfo {author} {\bibfnamefont {A.}~\bibnamefont
  {Widom}}, \bibinfo {author} {\bibfnamefont {J.}~\bibnamefont {Swain}},
  \bibinfo {author} {\bibfnamefont {J.}~\bibnamefont {Silverberg}}, \bibinfo
  {author} {\bibfnamefont {S.}~\bibnamefont {Sivasubramanian}}, \ and\ \bibinfo
  {author} {\bibfnamefont {Y.}~\bibnamefont {Srivastava}},\ }\href@noop {}
  {\bibfield  {journal} {\bibinfo  {journal} {Phys. Rev. E},\ }\textbf
  {\bibinfo {volume} {80}} (\bibinfo {year} {2009})}\BibitemShut {NoStop}%
\bibitem [{\citenamefont {Saija}\ \emph {et~al.}(2010)\citenamefont {Saija},
  \citenamefont {Aliotta}, \citenamefont {Fontanella}, \citenamefont
  {Pochylski}, \citenamefont {Salvato}, \citenamefont {Vasi},\ and\
  \citenamefont {Ponterio}}]{Sa10}%
  \BibitemOpen
  \bibfield  {author} {\bibinfo {author} {\bibfnamefont {F.}~\bibnamefont
  {Saija}}, \bibinfo {author} {\bibfnamefont {F.}~\bibnamefont {Aliotta}},
  \bibinfo {author} {\bibfnamefont {M.}~\bibnamefont {Fontanella}}, \bibinfo
  {author} {\bibfnamefont {M.}~\bibnamefont {Pochylski}}, \bibinfo {author}
  {\bibfnamefont {G.}~\bibnamefont {Salvato}}, \bibinfo {author} {\bibfnamefont
  {C.}~\bibnamefont {Vasi}}, \ and\ \bibinfo {author} {\bibfnamefont
  {R.}~\bibnamefont {Ponterio}},\ }\href@noop {} {\bibfield  {journal}
  {\bibinfo  {journal} {J. of Chem. Phys.},\ }\textbf {\bibinfo {volume}
  {133}},\ \bibinfo {pages} {081104} (\bibinfo {year} {2010})}\BibitemShut
  {NoStop}%
\end{thebibliography}
%

\end{document}